\documentclass[noshowpacs,amsmath,twocolumn,
aps,prb]
{revtex4-1}

\usepackage{setspace}
\usepackage{amsmath}
\usepackage{graphicx}
\usepackage[nearskip,margin = 0pt]{subfig}
\usepackage{subfig}

\usepackage{verbatim}
\usepackage{amsfonts}
\usepackage{amssymb}
\usepackage{epstopdf} 
\usepackage{xcolor}
\usepackage{color}
\DeclareGraphicsExtensions{.pdf,.eps,.png,.jpg,.mps} 
\usepackage{ragged2e}
\usepackage{hyperref}
\hypersetup{
    colorlinks=true,
    linkcolor=blue,
    citecolor=blue,
    filecolor=blue, 
    urlcolor=blue,
}
\usepackage{mathrsfs}

\usepackage{threeparttable}

\begin{document}

\title{Overcoming sensitivity-bandwidth trade-off in mid-infrared spectroscopy by a microresonator-anchored swept laser} 

\author{Zhaoyu Cai,$^{1}$ Zihao Wang,$^{1}$ Yulei Ding,$^{1}$ Yifei Wang,$^{1}$ Chengjiu Wang,$^{1}$ Changxi Yang,$^{1}$ 
Yanan Guo,$^{2,3}$ Jianchang Yan,$^{2,3}$ Junxi Wang,$^{2,3}$ Xun Liu,$^{4}$ Jiangtao Li,$^{4}$ Ruocan Zhao,$^{4,5}$ Xianghui Xue,$^{4,5,6,7}$ and Chengying Bao$^{1,*}$ \\
\vspace{1mm}
$^1$State Key Laboratory of Precision Measurement Technology and Instruments, Department of Precision Instruments, Tsinghua University, Beijing 100084, China.\\
$^2$College of Materials Science and Opto-Electronic Technology, University of Chinese Academy of Sciences, Beijing 100049, China.\\
$^3$Research and Development Center for Wide Bandgap Semiconductors, Institute of Semiconductors, Chinese Academy of Sciences, Beijing 100083, China.\\
$^4$CAS Key Laboratory of Geospace Environment, School of Earth and Space Sciences, University of Science and Technology of China, 230026 Hefei, China.\\
$^5$Hefei National Laboratory, University of Science and Technology of China, 230088 Hefei, China.\\
$^6$Hefei National Research Center for Physical Sciences at the Microscale and School of Physical Sciences, University of Science and Technology of China, 230026 Hefei, China.\\
$^7$CAS Center for Excellence in Comparative Planetology, Anhui Mengcheng Geophysics National Observation and Research Station, University of Science and Technology of China, 230026 Hefei, China.\\
Corresponding author:  $^*$cbao@tsinghua.edu.cn 
}

\maketitle
\newcommand{\ts}{\textsuperscript}

\newcommand{\tsb}{\textsubscript}


{\bf Optical frequency combs have revolutionized high-precision spectroscopy, yet an intrinsic trade-off between spectroscopic signal-to-noise ratio (sSNR) and measurement bandwidth ($B$) fundamentally constrains sensitive, broadband measurements. While broadband swept lasers offer a potential solution, generating broadband, ultrafast and linearly sweeping lasers with a narrow linewidth remains a significant challenge, particularly in the fingerprint mid-infrared (mid-IR) band. 
Here we overcome this limitation by using a microresonator-anchored ultrafast sweeping Fourier domain mode-locked (FDML) laser for mid-IR spectroscopy. We introduce a dual-microresonator-anchor approach: a microcomb provides frequency calibration and a high-Q microresonator resolves the instantaneous FDML lasing lineshape. The strategy enables accurate correction of the FDML laser's sweep nonlinearity and broad linewidth in the near-IR, allowing the FDML laser to function as a high-fidelity mid-IR light via difference frequency generation. The system achieves a record sSNR$\times$$B$ of 1.3$\times$10$^5$ THz$\cdot \sqrt{\rm Hz}$ and methane sensing precision of 9 ppb$\cdot$m$\cdot$$\sqrt{\rm s}$, while retaining GHz resolution to distinguish methane isotope. We further demonstrate broadband, coherent swept laser phase spectroscopy in the mid-IR, tolerating losses up to 78 dB. This work leverages advances in integrated photonics to overcome the fundamental limitations of precision spectroscopy, paving the way for next-generation, broadband, and ultra-sensitive mid-IR spectroscopic sensing systems.}

\begin{figure*}[t]
\begin{centering}
\includegraphics[width=0.98\linewidth]{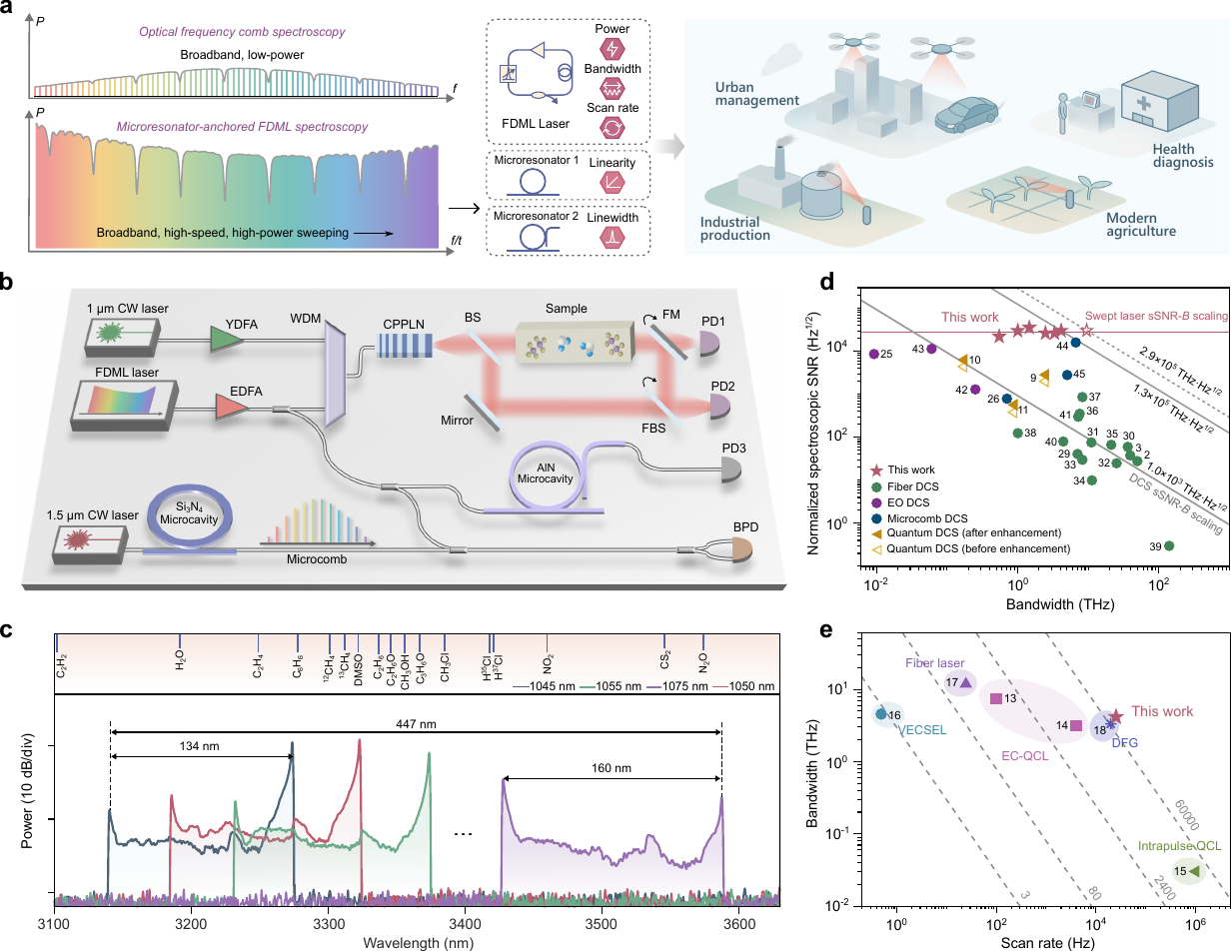}
\captionsetup{singlelinecheck=off, justification = RaggedRight}
\caption{\textbf{Concept and performance of microresonator-anchored mid-IR FDML spectroscopy}. \textbf{a,} Microresonator-anchored DFG FDML laser can be used for precise mid-IR spectroscopy, breaking the universal trade-off between sSNR and $B$ for comb spectroscopy. With dual-microresonator-calibration, the FDML laser features high power, broad bandwidth and fast scan rate, as well as high effective high sweep linearity and narrow linewidth, making it invaluable for many applications. 
\textbf{b,} Experimental setup for microresonator-anchored mid-IR FDML spectroscopy. An integrated Si$_3$N$_4$ soliton microcomb calibrates the FDML lasing frequency, while an AlN microresonator with a drop-port measures the lasing lineshape. FBS, flip beam splitter; FM, flip mirror; BPD, balanced photodetector.   
\textbf{c}, Mid-IR FDML laser spectra generated via DFG. Numerous gas samples are covered within the available lasing bandwidth. 
\textbf{d}, Mid-IR FDML spectroscopy performance compared with DCS (including near-IR DCS) reports. Open star means measurement stitching three 1 $\mu$m pump wavelengths. 
\textbf{e}, Mid-IR FDML laser performance compared with other mid-IR swept lasers, showing an excellent combination of bandwidth and scan rate.
} 
\label{Fig1}
\end{centering}
\end{figure*}

Broadband, comb-based mid-infrared (mid-IR) spectroscopy enables sensitive, parallel detection of multi-species trace gases, impacting modern society activities in environmental protection, health diagnosis and agricultural productions \cite{Hansch_NP2012mid,Newbury_NP2018high,Vodopyanov_NP2018massively,Ye_Nature2025modulated,Picque_NP2019frequency}. 
In particular, dual-comb spectroscopy (DCS), which leverages the relative scan between two combs with a slightly different repetition rate for effective Fourier transform spectroscopy, has become a field-deployable technique for measurements of trace gases  \cite{Newbury_Optica2016,Picque_NP2019frequency,Newbury_Optica2019mid,Pan_NP2024dual}. DCS is free from moving parts and renowned for its fast speed and high spectral resolution \cite{Newbury_Optica2016}. However, it faces a critical trade-off between spectroscopic signal-to-noise ratio (sSNR) and measurement bandwidth ($B$), as a large comb line number $M$ limits the power per line. This trade-off is universal for comb-based spectroscopy. sSNR is crucial as it determines both the achievable sensitivity and the measurement time needed to reach a certain precision \cite{Picque_NP2019frequency}. Although $M\times$sSNR has been used as a metric to characterize DCS systems \cite{Newbury_Optica2016} (Fig. \ref{Fig1}), the absorbance of gas samples is usually determined by a few lines around a narrow absorption branch. Therefore, enhancing the sSNR is of fundamental interests to upgrade DCS systems. Squeezed or entangled DCS has been reported to enhance sSNR \cite{Diddams_Science2025squeezed,Zhang_PRX2025entangled,Zeng_LSA2025quantum}. Nevertheless, the enhancement is limited to a few dB.

In this work, we overcome the trade-off between sSNR and $B$ by using a Fourier domain mode-locked (FDML) laser for fast, broadband, mid-IR wavelength swept laser spectroscopy \cite{goldenstein2017infrared} (Fig. \ref{Fig1}a). Wavelength swept lasers can provide higher lasing power than individual comb lines, which is not impacted by the bandwidth. Thus, it can break the above sSNR-$B$ trade-off and improve sSNR. However, wavelength swept lasers encompass fast sweep rates, broad bandwidths, high frequency swept linearity and narrow linewidth remain a significant challenge, especially in the mid-IR \cite{strand2019measurement,revin2022fast,chrystie2014ultra,rey2014broadly,woodward2019swept,abe2016rapid}. FDML laser, which was first developed for optical coherent tomography (OCT), provides ultrahigh chirp rate and broad bandwidth in the near-infrared (near-IR) \cite{Fujimoto_OE2006fourier}. Recently, we have shown that its ultrafast frequency sweep can be precisely calibrated by a large line spacing photonic-chip microcomb \cite{Kippenberg_Science2018Review,Bowers_NP2022integrated,Bao_SA2025microcomb}. Converting it into the mid-IR by difference frequency generation (DFG) can resolve the above challenge for fast and linearly sweeping, broadband mid-IR lasers via near-IR calibration. However, application of FDML lasers in spectroscopy has been limited to samples with broad absorption signatures (e.g., gases at high pressure and temperature) due to its broad instantaneous linewidth \cite{kranendonk2007high,ma201350}. Here, we use another high-Q microresonator with a drop-port to measure the instantaneous lineshape and to correct its influence on spectroscopy.

With the dual-microresonator-anchor in the near-IR, the DFG FDML laser achieves GHz resolution, enabling isotope differentiation and a metric of sSNR$\times B$ as high as 1.3$\times$10$^5$ THz$\cdot \sqrt{\rm Hz}$ in the mid-IR (two orders of magnitude higher than typical DCS systems, Fig. \ref{Fig1}). The high sSNR also enables a normalized precision of 9 ppb$\cdot$m$\cdot$$\sqrt{\rm s}$ for methane detection, which is orders of magnitude higher than mid-IR DCS results \cite{jerez2018flexible,Vahala_NC2021architecture}. The dual-calibrated FDML laser was used for broadband, coherent phase spectroscopy in the mid-IR for the first time, to our knowledge. This coherent measurement shows a capability to tolerate loss up to 78 dB for spectroscopy with 20 ms averaging, which is advantageous in high-loss scenarios involving non-cooperative targets. Our work constitutes a promising advance in broadband, high sSNR spectroscopy as well as tunable mid-IR lasing sources.

\begin{figure*}[t!]
\begin{centering}
\includegraphics[width=0.98\linewidth]{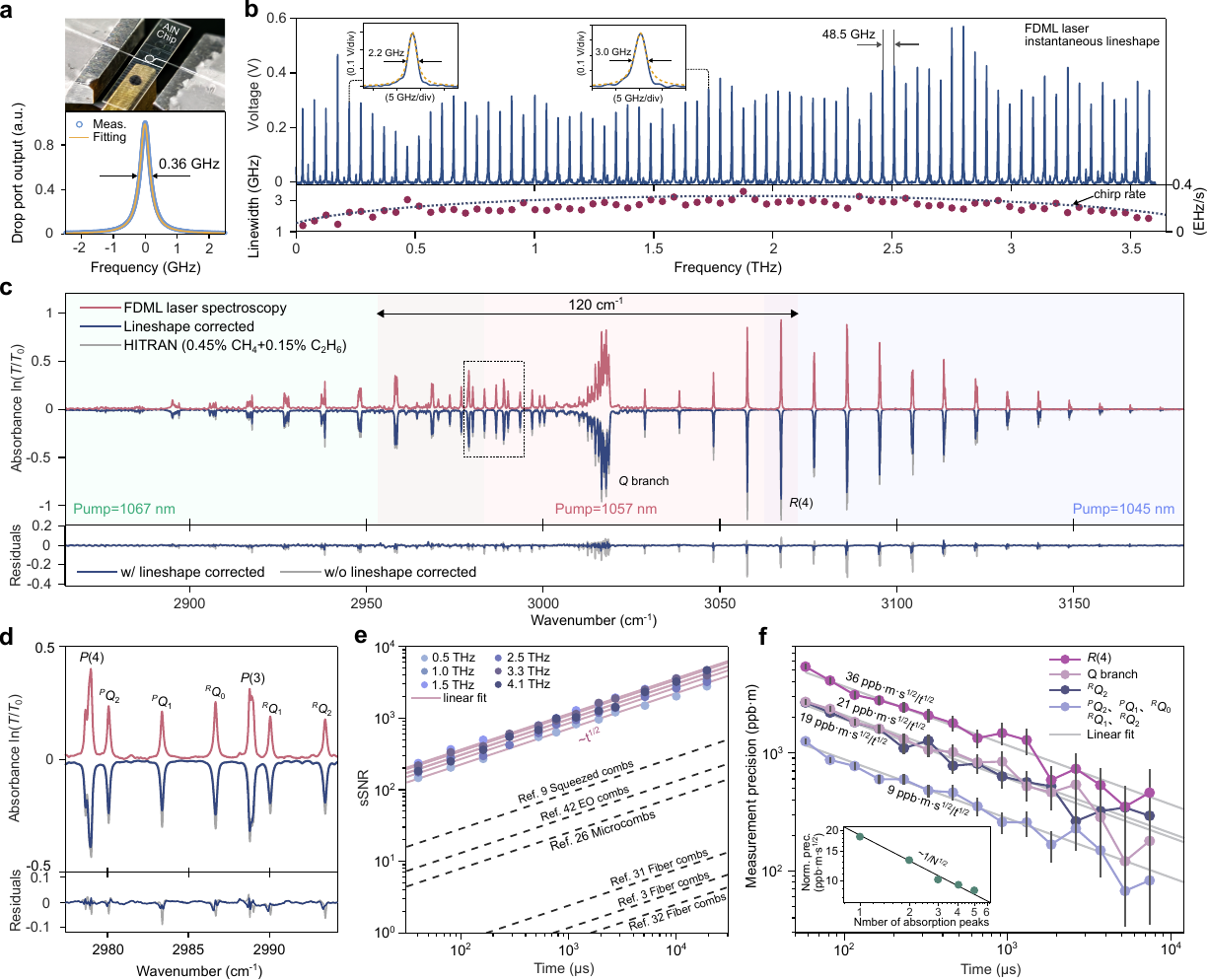}
\captionsetup{singlelinecheck=off, justification = RaggedRight}
\caption{\textbf{FDML laser lineshape measurement and mid-IR spectroscopy}. \textbf{a}, AlN chip and transmission measured at its drop-port. The microresonator has a linewidth of 0.36 GHz based on a Lorentzian fit.
\textbf{b}, Measured FDML laser lineshape and the corresponding linewidth. The inset shows zoom in of the measured lineshape. 
The bottom panel shows the linewidth of the FDML laser, which generally follows the chirp rate of the FDML laser (dashed curve). 
\textbf{c}, Measured absorption spectrum of a gas mixture of methane and ethane spanning 9.5 THz by stitching three 1.06 $\mu$m CW pump wavelengths. 
The gray and blue curves show the reference spectra without and with accounting for the FDML laser lineshape. 
The bottom panel shows the residual error between the measured result and the reference spectra with or without lineshape correction.
\textbf{d}, Zoom in of the dashed box in panel \textbf{c}.  
\textbf{e}, sSNR under different FDML lasing bandwidths, all scaling as $\sqrt{t}$ ($t$ is the measurement time) and showing similar sSNR. The dashed lines are sSNR for reported DCS systems. 
\textbf{f}, Measurement precision of methane and ethane concentration evaluated by Allan deviation using different rovibrational branches annotated in panels \textbf{c, d}. The inset shows the measurement precision can be enhanced by using multi-branches for fit. The normalized precision improves as 1/$\sqrt{N}$ ($N$ is the number of used branches).} 
\label{Fig2}
\end{centering}
\end{figure*}

\begin{figure}[t!]
\begin{centering}
\includegraphics[width=0.98\linewidth]{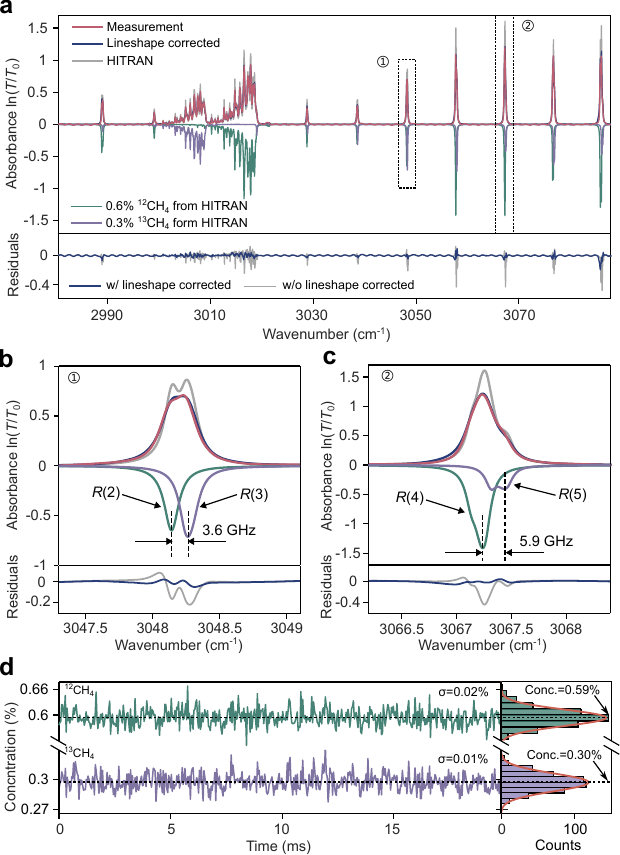}
\captionsetup{singlelinecheck=off, justification = RaggedRight}
\caption{\textbf{Mathane isotope spectroscopy.} \textbf{a}, Absorption spectrum measured by the DFG FDML laser and reference spectra with and without considering laser lineshape correction. The residual error greatly reduces by including lineshape correction (bottom panel).
\textbf{b, c}, Zoom in of the boxed absorption branches in panel \textbf{a}. Absorption branches from $^{12}$CH$_4$ and $^{13}$CH$_4$ overlap in this measured band. The measured absorption spectra only agree with the reference spectra when accounting for the FDML laser lineshape. 
\textbf{d}, Gas concentrations of $^{12}$CH$_4$ and $^{13}$CH$_4$ derived by using the absorption spectra in \textbf{b}. The right panels show the distribution of the gas concentration.
} 
\label{Fig3}
\end{centering}
\end{figure}

\begin{figure*}[t!]
\begin{centering}
\includegraphics[width=0.98\linewidth]{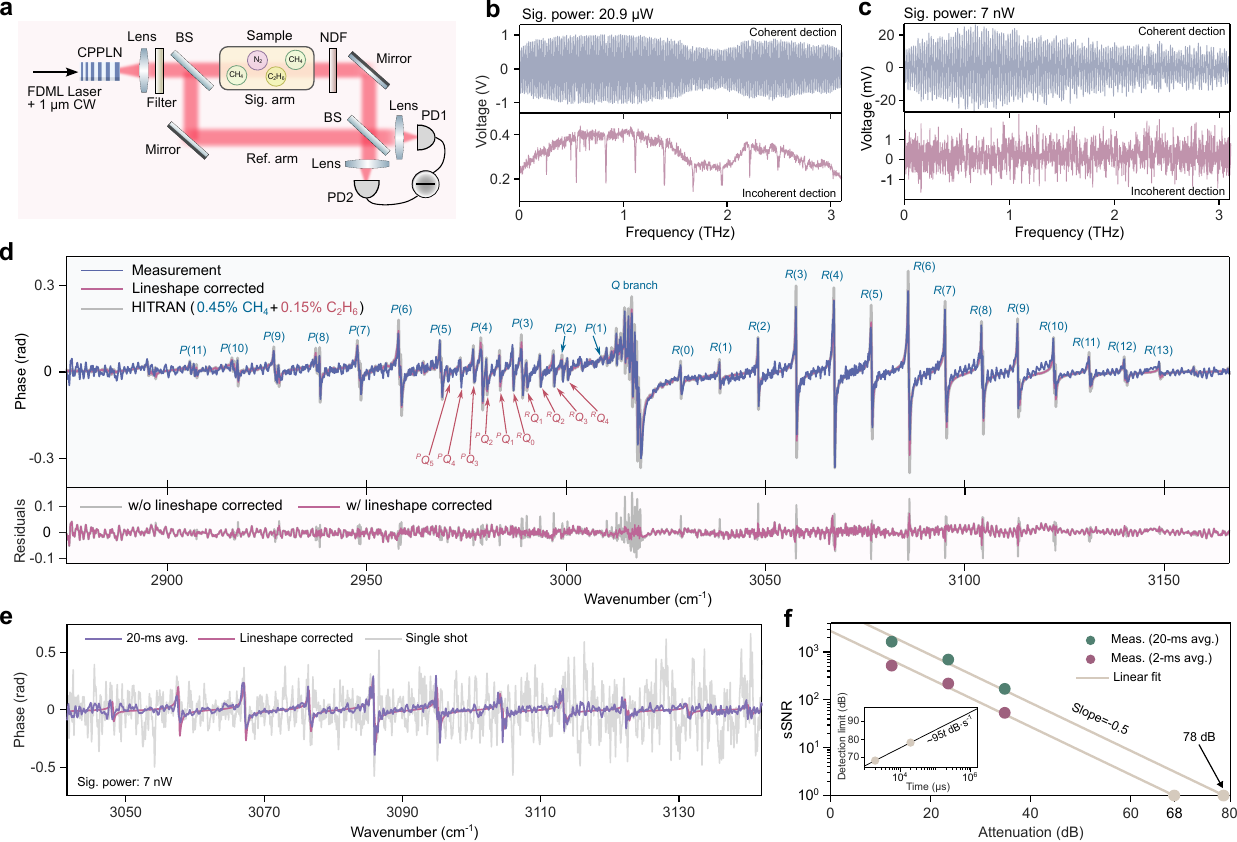}
\captionsetup{singlelinecheck=off, justification = RaggedRight}
\caption{\textbf{Coherent phase spectroscopy.} 
\textbf{a}, Experimental setup for coherent phase spectroscopy using the microresonator-anchored FDML laser. Neutral density filters (NDFs) with different attenuation levels were inserted to evaluate the system under different received signal powers. PD, photodetector; BS, beam splitter.
\textbf{b, c}, Measured signals using coherent or incoherent detection scheme. The labeled signal powers are total signal power on both detectors.
\textbf{d}, Measured phase spectroscopy over 9 THz range by stitching three 1.06 $\mu$m pump wavelengths; the bottom panel shows the residual error with and without including the lineshape correction.
\textbf{e}, Phase spectra measured by a single shot and averaged over 20 ms, with a signal arm power of 7 nW and a pump wavelength of 1048 nm. 
\textbf{f}, sSNR of the absorption phase spectra measured (averaged in 2 or 20 ms) under different NDF attenuation factors. The inset shows the tolerable loss versus averaging time when taking sSNR=1 as the detection limit.  
}
\label{Fig4}
\end{centering}
\end{figure*}


\noindent \textbf{Concept of dual-anchored mid-IR FDML spectrometer.} Our mid-IR FDML spectrometer is illustrated in Figs. \ref{Fig1}a, b. Wavelength sweep in the FDML laser relies upon a fibre Fabry-Perot tunable filter (FFP-TF). When the modulation frequency of the FFP-TF matches the free spectral range (FSR) of the laser cavity, FDML can be established to enable broad sweeping bandwidths and ultra-high chirp rates \cite{Fujimoto_OE2006fourier,Bao_SA2025microcomb}. Our FDML laser can have a bandwidth of 4.2 THz and a modulation frequency of 24.6 kHz. To attain ultrafast wavelength swept mid-IR emission, we combine the 1.55 $\mu$m FDML laser with a 1.06 $\mu$m continuous wave (CW) laser in a chirped periodically poled lithium niobate (CPPLN) waveguide for DFG (Methods and Supplementary Fig. S1). The CPPLN design enables broadband phase-matching and conversion into the mid-IR with a spectrum spanning 160 nm (Fig. \ref{Fig1}c). As the CPPLN also has a broad phase matching bandwidth at 1.06 $\mu$m, the centre wavelength of the mid-IR output can be further tuned by about 450 nm through adjustment of the wavelength of the 1.06 $\mu$m CW laser, covering various gas samples (Fig. \ref{Fig1}c).

As the FFP-TF has a bandwidth of 7.5 GHz, the FDML laser should have a relatively broad instantaneous bandwidth. It remains a significant challenge to measure the lasing lineshape (not mere bandwidth) over the entire FDML lasing bandwidth in real time. Previous measurements of FDML laser linewidth rely upon picking a snapshot of the broad spectrum to be measured by a slow optical spectrum analyzer \cite{biedermann2010direct}. Here, we used an AlN microresonator with a drop-port \cite{Bao_OL2021self} and resonance linewidths narrower than the potential FDML laser linewidth to measure the laser lineshape over the entire bandwidth in real time. It allows correction of the influence of broad linewidth on precise spectroscopy (see next section). In addition, a 50 GHz Si$_3$N$_4$ microcomb was used to calibrate the instantaneous lasing frequency \cite{Bao_SA2025microcomb}. 

With the dual-microresonator-calibration, the mid-IR FDML laser overcomes the sSNR-$B$ trade-off for precise spectroscopy (Fig. \ref{Fig1}d). Owing to this fundamental constraint, most DCS systems cluster around an sSNR$\times B$=1.0$\times$10$^3$ THz$\cdot \sqrt{\rm Hz}$ \cite{
liu2023mid,muraviev2020broadband,lind2020mid,guo2020nanophotonic,chen2019mid,nader2019infrared,timmers2018molecular,Vodopyanov_NP2018massively,Newbury_NP2018high,zhu2015mid,abbas2019time,baumann2011spectroscopy,okubo2015ultra,hoghooghi2019broadband,bernhardt2010cavity,jerez2018flexible,yan2017mid,millot2016frequency,Vahala_NC2021architecture,yu2018silicon,wang2025rhythmic,Diddams_Science2025squeezed,Zeng_LSA2025quantum,Zhang_PRX2025entangled} (gray line in Fig. \ref{Fig1}d), despite wide variations in reported systems. In sharp contrast, sSNR of our FDML spectrometer remains independent from $B$ (red line in Fig. \ref{Fig1}d).

As noted above, mid-IR wavelength swept lasers can hardly encompass large bandwidth, fast sweep rate and narrow lasing linewidth (Fig. \ref{Fig1}e and Supplementary Table S2). For example, wavelength swept external cavity quantum cascaded lasers (EC-QCLs) can have relatively broad bandwidth, but suffer from low sweep rate (limited to a few kHz) and mode-hopping\cite{strand2019measurement,revin2022fast}. Intrapulse QCLs can achieve MHz tuning rate but only have a sweep bandwidth limited to a few cm$^{-1}$ \cite{chrystie2014ultra}. Although ref. \cite{abe2016rapid} reports similar bandwidth and scan rate, it has a much broader linewidth (Supplementary Table S2). 
Our work resolves a longstanding challenge in realizing broadband, fast-swept mid-infrared lasers with effectively narrow linewidth and high sweep linearity (Fig. \ref{Fig1}a).


\noindent \textbf{High sSNR mid-IR FDML spectroscopy.} In the experiment, a 49 GHz AlN microresonator was used to measure the instantaneous lineshape of the FDML laser. The transmission in the drop-port has a linewidth of 0.36 GHz (Fig. \ref{Fig2}a), sufficient to resolve the laser lineshape. Note that a too high Q-factor (very narrow linewidth) may be inappropriate for this lineshape measurement, as it reduces the transmitted power. The narrow microresonator linewidth samples the instantaneous FDML lasing spectrum, and the measured laser lineshapes are plotted in Fig. \ref{Fig2}b. When zooming in the lineshape, it generally exhibits a Lorentzian shape, but is asymmetric (inset of Fig. \ref{Fig2}b). The linewidth ranges from 1 to 3 GHz, smaller than the 7.5 GHz FFP-TF bandwidth. The linewidth tends to be narrower at the edge of the lasing span, which could result from the reduced chirp rate at the edge of the sweep span (see dashed curve in the bottom panel of Fig. \ref{Fig2}b). These FDML lasing dynamics are first observed here, to our knowledge. Together with instantaneous frequency measured by a microcomb \cite{Bao_SA2025microcomb}, high-Q microresonators constitute powerful tools to characterize FDML dynamics to understand its lasing physics in the future. As an aside, both wavelength-dependent drop-port transmission and lasing power variation can contribute to the observed power variation in Fig. \ref{Fig2}b.

With the lasing frequency and lineshape calibrated in the near-IR, we used the DFG FDML laser for mid-IR spectroscopy of a 5 cm gas cell, filled with 0.45\% methane (CH$_4$) and 0.15\% ethane (C$_2$H$_6$) and buffered by nitrogen to a total pressure of 760 Torr to mimic an ambient condition (concentrations can have a variation of $\pm2\%$). When setting the 1.06 $\mu$m CW laser at 1057 nm, the measurement can span 3.6 THz (limited by the microcomb bandwidth). By further tuning the 1.06 $\mu$m laser to 1067 and 1045 nm, a spectroscopy bandwidth of 9.5 THz can be realized (red curve in \ref{Fig2}c). Since the FDML laser linewidth is comparable to the absorption bandwidth, the direct measured absorbance tend to be broader and weaker than the HITRAN database. The relatively large residual error between them is shown by the gray curve in the bottom panel of Fig. \ref{Fig2}c. Then, we corrected the absorption spectrum for the gases using the measured laser lineshape (Methods). The measurement agrees well with the corrected spectrum (blue curve in the bottom panel of Fig. \ref{Fig2}c). To better visualize the agreement, we zoom in the dashed boxed region in Fig. \ref{Fig2}d. The retrieved concentrations of CH$_4$ and C$_2$H$_6$ are 0.45$\%$ and 0.15$\%$, respectively, which agree well with the set values of the cell. However, these concentrations differ from the cell, if the linewidth correction is not accounted for (Supplementary Fig. S2).

Then, we characterize the sSNR of the system (Methods), which scales as $\sqrt{t}$ ($t$ is the measurement time). We adjusted the bandwidth of the FDML laser by controlling the drive power of the FFP-TF (Supplementary Fig. S3), and did not observe evident change of sSNR with the FDML sweep bandwidth (Fig. \ref{Fig2}e). The highest normalized sSNR reaches 3.6$\times$10$^4$ $\sqrt{\rm Hz}$, much higher than typical DCS systems; note that ref. \cite{yu2018silicon} is not suitable for gas spectroscopy due to its low, 127 GHz resolution. The reference arm in Fig. \ref{Fig1}b is essential to achieve this sSNR by canceling the intensity noise (Methods and Supplementary Fig. S4). 
Given a measurement bandwidth $B$=3.6 THz, it yields sSNR$\times$$B$=1.3$\times$10$^5$ THz$\cdot$$\sqrt{\rm Hz}$. When stitching three 1.06 $\mu$m pump wavelengths, sSNR$\times$$B$ can be further boosted to a 2.9$\times$10$^5$ THz$\cdot$$\sqrt{\rm Hz}$ (open star in Fig. \ref{Fig1}d). 

The high sSNR secures high measurement precision. We evaluated the Allan deviation of the retrieved concentrations of CH$_4$ and C$_2$H$_6$ in Fig. \ref{Fig2}f, scaling as 1/$\sqrt{t}$. The normalized precision for CH$_4$ using the $Q$ branch or the $^RQ$2 branch of C$_2$H$_6$ for concentration retrieval reaches 20 ppb$\cdot$m$\cdot$$\sqrt{\rm s}$, while the normalized precision retrieved from the $R$(4) branch of CH$_4$ is 36 ppb$\cdot$m$\cdot$$\sqrt{\rm s}$. 
There are five absorption branches around 2985 cm$^{-1}$ exhibit similar absorbance for C$_2$H$_6$. When using all the five branches for fitting to derive the concentration, the normalized precision improves to 9 ppb$\cdot$$\rm m$$\cdot \rm \sqrt{s}$. The inset of Fig. \ref{Fig2}f confirms the measurement precision improves as 1$/\sqrt{N}$, where $N$ is the number of branches used for fitting. The precision is 2 orders of magnitude higher than using microcomb as light source (as opposed to as a frequency ruler here) for mid-IR DCS of methane \cite{Vahala_NC2021architecture}. It also suggests that broad measurement bandwidth not only enables multi-species gases detection, but also enhances the measurement precision of a single sample by discerning more absorption branches. 

As an aside, we used an oscilloscope with a bandwidth of 8 GHz for frequency calibration (losing calibration signal when the beat frequency between the FDML laser and the microcomb is higher than 8 GHz). No evident spectroscopy precision improvement was observed when switching the digitizer bandwidth to 24 GHz, while ranging precision of an FDML LIDAR increases with the full sweep span calibrated \cite{Bao_SA2025microcomb} (Supplementary Fig. S5).

\noindent \textbf{High resolution spectroscopy of methane isotope.} To show high resolution in the mid-IR spectroscopy, we exploited the dual-calibrated DFG FDML laser for spectroscopy of another 4.5 cm gas cell comprising methane isotope (0.6\% $^{12}$CH$_4$ and 0.3\% $^{13}$CH$_4$). This cell was also buffered to an ambient pressure of 760 Torr by nitrogen. The measured spectrum of the gas cell is plotted in Fig. \ref{Fig3}a (1.06 $\mu$m pump laser set at 1055 nm).  In general, the absorption spectra of $^{13}$CH$_4$ are red-shifted by $\sim $10 cm$^{-1}$ from $^{12}$CH$_4$, and some of their absorption branches in the $\nu_3$ vibrational absorption band partially overlap. By considering the lineshape correction, the residual error between the measurement and the corrected absorption spectrum is minimized. The oscillatory background in the residual error can result from FP effect of the gas cell, which should be removed by optimal cell design and fabrication.

Here, we focused on the $R$ branches (dashed boxed regions of Fig. \ref{Fig3}a), where the isotope absorption branches overlap, to evaluate the influence of lineshape correction and spectral resolution of our system. 
Due to the relatively broad FDML laser linewidth, the measured spectra differ from the mixed isotope absorption spectra (gray curves in Figs. \ref{Fig3}b, c). Equivalently, the uncalibrated FDML laser cannot resolve the isotope. The discrepancy is corrected when considering the near-IR FDML lineshape, see blue curves and residual error in the bottom panels of Figs. \ref{Fig3}b, c. Based on the correction, we derived the $^{12}$CH$_4$ and $^{13}$CH$_4$ concentrations using the $R$(2) and $R$(3) branches, respectively. The concentrations are 0.59$\pm$0.02\% for $^{12}$CH$_4$ and 0.30$\pm$0.01\% for $^{13}$CH$_4$ with a single measurement time of 41 $\mu$s (Fig. \ref{Fig3}d). These concentrations are consistent with the preset values of the gas cell. Moreover, Allan deviation of the concentrations is similar to Fig. \ref{Fig2}f (Supplementary Fig. S6), which suggests the measurement precision is not impacted by the overlapped absorption spectrum and our system should have a resolution higher 3 GHz.

\noindent \textbf{Coherent mid-IR swept laser spectroscopy.} We further used the calibrated FDML laser for phase coherent spectroscopy (Fig. \ref{Fig4}a). Leveraging a strong local oscillator in the reference arm, such a scheme can be advantageous in spectroscopy with a weak received signal. However, broadband, phase coherent swept laser spectroscopy has not been demonstrated in the mid-IR, to our knowledge. In the experiment, we delayed the signal and reference arm by about 2 cm, and the interference signal was measured by balanced photodetection (Methods). An example of the measured interferogram is shown in Fig. \ref{Fig4}b. Similar to incoherent measurement, absorption dips can be observed in the interferogram. Evident interferogram can still be observed when attenuating the received signal by 35 dB to 7 nW (Fig. \ref{Fig4}c). In contrast, no clear absorption dip can be observed for incoherent measurement with such a low received power (see also Supplementary Sec. 8). 

The phase spectrum can be retrieved using the measured interferogram (Methods). When stitching measurements using three 1.06 $\mu$m CW pump wavelengths, phase spectrum over 9 THz can be measured (Fig. \ref{Fig4}d). The received signal power was 20.9 $\mu$W, while the averaging time was 20 ms for this measurement. The measured absorption phase agrees with the gas absorption, when considering the lineshape correction (see blue and red curves). Without this correction, the measured phase differs from the gas absorption, see particularly the $Q$ branch of methane and the gray curve for the residual error (bottom panel of Fig. \ref{Fig4}d). 

When attenuating the received signal power to 7 nW, the phase spectrum measured in a single shot is strongly perturbed by noise (gray curve in Fig. \ref{Fig4}e). By averaging the signal over 20 ms, phase spectrum in agreement with the corrected gas absorption phase can be attained (blue and red curves). Even with this 35 dB attenuation, the phase spectrum reached an sSNR of 170 after 20 ms averaging (Methods). We also evaluated sSNR of the phase spectra under attenuation of 12.2 dB and 23.5 dB. The sSNR measured in 2 ms or 20 ms decreases an inverse-square-root scaling way with attenuation (Fig. \ref{Fig4}f), while it decreases inversely with attenuation in incoherent detection (Supplementary Fig. S8). When taking the fitted sSNR=1 as the detection limit, our system tolerates 78 (68) dB loss in 20 (2) ms measurement time (35 dB higher than incoherent detection). In principle, the tolerated loss further increases as 95$t$ dB/s (inset of Fig. \ref{Fig4}f). These measurements showcase the effectiveness of coherent FDML mid-IR spectroscopy in high-loss scenarios, including reflections from non-cooperative targets and transmission through scattering media such as flames or fogs \cite{mitchell_Optica2018coherent}. 

\noindent \textbf{Discussion.} We have demonstrated a viable approach to overcome the intrinsic trade-off in sSNR and $B$ for comb-based spectroscopy. As opposed to using the comb as a light source for spectroscopy, we used an integrated microcomb and a passive microresonator to anchor an ultrafast sweeping FDML laser for high-sSNR, high-resolution, broadband mid-IR spectroscopy. Our system exhibits a record sSNR$\times B$, while retaining GHz-resolution and 25 kHz acquisition rate (see comprehensive comparision with DCS reports in  Supplementary Table S1). The sSNR enhancement is also much higher than quantum-enhanced DCS \cite{Diddams_Science2025squeezed,Zhang_PRX2025entangled,Zeng_LSA2025quantum}. 
FDML lasers can be optimized toward tens of THz bandwidth and MHz modulation rate \cite{Huber_BOE2017high}, which can further elevate sSNR$\times B$ and measurement speed. A broader microcomb will be needed to pair with such a broader FDML laser. Tight mode confinement and dispersion engineering of nanophotonic devices can enable low dispersion to support such microcombs \cite{Kippenberg_Science2018Review}. Dispersion engineering of thin-film PPLN waveguides can also allow broad DFG phase matching bandwidth without using the chirped design \cite{Fejer_OE2022ultra}. The uniform poling period and tight mode confinement can improve the DFG efficiency, leading to energy-efficient spectroscopic sensors. 
Multiple waveguides with different poling designs can be accommodated in a single photonic chip to target various mid-IR wavelengths. For instance, 4$-$5 $\mu$m emissions can be accessed by appropriate poling design and shifting the CW pumping wavelength to around 1.15 $\mu$m. Thus, a microresonator-anchored FDML laser can be distributed across multiple PPLN waveguides for DFG, enabling parallel interrogation of various samples via either incoherent or coherent spectroscopy. Such a next-generation spectroscopic system can be used for detecting gas leak\cite{Newbury_Optica2019mid}, identifying super-emitter of methane \cite{duren_Nature2019california}, and health diagnose \cite{Ye_Nature2025modulated} (Fig. \ref{Fig1}a).

\vspace{3 mm}
\noindent{\bf Methods}

{\small
\noindent \textbf{FDML laser, Si$_3$N$_4$ microcomb, AlN microresonator and CPPLN.}  Key parameters of the FDML laser and the Si$_3$N$_4$ microcomb are summarized as follows (further details can be found in ref. \cite{Bao_SA2025microcomb}). The FDML laser cavity comprised 7.2 km single mode fibre and 0.9 km dispersion compensating fibre. It has an output power of 1.3 mW. 
The foundry manufactured Si$_3$N$_4$ microresonator has an intrinsic and loaded Q-factors of 9.8 million and 6.4 million, respectively. 
We pumped the microresonator by a CW laser with an on-chip power of 200 mW to generate a soliton microcomb with a 3 dB bandwidth of about 24 nm. 

The AlN microresonator has a diameter of 900 $\mu$m and was fabricated on the AlN-on-sapphire platform \cite{Bao_OL2021self}. A 0.9 $\mu$m thick crystalline AlN film was epitaxially grown by metal–organic chemical vapor deposition (MOCVD) on a sapphire substrate. The film was etched to a waveguide dimension of 2.2$\times$0.9 $\mu$m using the same fabrication method reported in ref. \cite{Bao_OL2021self}. It has an intrinsic and loaded Q-factors of 0.8 million and 0.55 million, respectively. 

The CPPLN is designed with a linearly varying poling period. Due to the chirped poling design, it allows a phase matching bandwidth of 37 nm at 1550 nm. The CPPLN has an efficiency of 0.13 \%/W. The optical power was set to 160 mW at 1.55 $\mu$m and 650 mW at 1.06 $\mu$m, yielding an estimated output power of 130 $\mu$W in the mid-IR. Although the chirping is not designed for the 1.06 $\mu$m band, the CPPLN naturally has a phase matching bandwidth as broad as 58 nm at 1.06 $\mu$m (Supplementary Fig. S1). Hence, we can shift the operation band by tuning the 1.06 $\mu$m laser, see Fig. \ref{Fig1}c.

\vspace{1 mm}

\noindent \textbf{Intensity reference and sSNR.} FDML lasers typically have considerable intensity noise. To minimize its influence, we used PD2 in Fig. \ref{Fig1}b to monitor the intensity fluctuations. Due to the unbalanced optical and electronic paths, there is a relative time delay  between the signal and reference arms. This delay was derived via cross-correlation, which was used to align the measured signal and reference traces temporally. The absorption was derived by normalizing the signal trace ($T$) using the reference trace ($T_0$). This reference contributes to nearly an order-of-magnitude enhancement in sSNR (Supplementary Sec. 4).

Following ref. \cite{Newbury_OE2010sensitivity}, the sSNR for magnitude spectra in Fig. \ref{Fig2} is defined as,

\begin{equation}
\begin{aligned}
{\rm sSNR}(\nu)\equiv\frac{1}{\sigma(\nu)},
\end{aligned}
\label{eq0sSNR}
\end{equation}
where $\sigma(\nu)$ is the standard deviation of the random fluctuations in the non-absorbing regions of the magnitude spectra.  sSNR for phase spectroscopy in Fig. \ref{Fig4} is defined in a similar way by setting $\sigma(\nu)$ to quantify the fluctuations in the measured phase spectra.

\vspace{1 mm}

\noindent \textbf{Laser lineshape correction in spectroscopy.} The measured FDML lasing lineshape $g_{\rm L}(\nu)$ allows us to correct its influence on spectroscopy. 
With the broaden lasing linewidth, transmission spectrum of a gas sample can be written as,
\begin{equation}
\begin{aligned}
T_{\rm L}(\nu)= & \frac{ \int T_{\rm H}(\nu')g_{\rm L}(\nu'-\nu){\rm d}\nu'}{\int g_{\rm L}(\nu'-\nu){\rm d}\nu'}.
\end{aligned}
\label{eq2Correction}
\end{equation}
where $T_{\rm H}(\nu')$ is the sample absorption in the HITRAN database. In our experiments, we segmented the lineshape measurements (Fig. \ref{Fig2}b) over the whole FDML lasing bandwidth by an interval of 49 GHz. Each lineshape is assigned to the corresponding $T_{\rm H}(\nu')$ to calculate the resulting $T_{\rm L}(\nu)$, which is shown as the blue curve in Fig. \ref{Fig2}c. The absorbance with lineshape correction was then derived as $A_{\rm L}(\nu)=-{\rm ln}[T_{\rm L}(\nu)]$. Fitting the measured absorbance to $A_{\rm L}(\nu)$ yields the gas concentration (Supplementary Sec. 2).

\vspace{1 mm}

\noindent \textbf{Coherent phase spectroscopy.} The experimental setup for coherent phase spectroscopy is illustrated in Fig. \ref{Fig4}a. A Mach–Zehnder interferometer was used for coherent detection. The gas cell is inserted into one arm (the signal arm) of the interferometer, while the other acts as the reference arm. The two arms were delayed by about 2 cm, and the resulting interference signals were detected by two mid-IR detectors (Vigo, PVI-4TE-4). The $\phi$ delay between two detected signals enables balanced detection to enhance the sSNR. The power in the reference arm (i.e., local oscillator) was 17 $\mu$W. 

The measured data have time as the x-axis; we resampled the data based on the instantaneous frequency calibration. The resampled data have instantantaneous frequency $\nu_s$ as the x-axis (see more details in ref. \cite{Bao_SA2025microcomb}). 
Then, the normalized, resampled interference signal can be written as,
\begin{equation}
\begin{aligned}
V_{b}(\nu_s) ={\rm cos}\left[2\pi \Delta t \nu_s+\phi_{\rm M}(\nu_s)\right],
\end{aligned}
\label{eq1}
\end{equation}
where $\Delta t$ is the temporal delay between two arms, and $\phi_{\rm M}(\nu_s)$ is the phase change induced by absorption in the signal path. To retrieve the absorption-induced phase change, we Hilbert transformed the resampled data $V_{b}(\nu_s)$ to derive its instantaneous phase $2\pi \Delta t \nu_s+\phi_{\rm M}(\nu_s)$. Then, we subtracted the linear phase change $2\pi \Delta t$ to have the measured $\phi_{\rm M}(\nu_s)$ (blue curve in Fig. \ref{Fig4}d). 

Similar to Eq. \ref{eq2Correction}, the complex absorption spectrum $\widetilde{T_{\rm L}}(\nu)$ accounting for the finite FDML lasing linewidth can be written as, 
\begin{equation}
\begin{aligned}
\widetilde{T_{\rm L}}(\nu)= & \frac{ \int T_{\rm H}(\nu')e^{i\phi_{\rm H}(\nu')}g_{\rm L}(\nu'-\nu){\rm d}\nu'}{\int g_{\rm L}(\nu'-\nu){\rm d}\nu'}.
\end{aligned}
\label{eq1}
\end{equation}
where $\phi_{\rm H}(\nu')$ is the phase spectrum derived from the HITRAN database using the Kramers-Kronig relationship. The complex absorption spectrum can be further written as $\widetilde{T_{\rm L}}(\nu)=T_{\rm L}(\nu)e^{i\phi_{\rm L}(\nu)}$, where $\phi_{\rm L}(\nu)$ is the lineshape-corrected phase spectrum (i.e., red curve in Fig. \ref{Fig4}d).





}

\vspace{3 mm}
{\small

\noindent \textbf{Acknowledgements.}
We thank Prof. Qiang Liu and Prof. Yidong Tan at Tsinghua University for discussions and equipment loan. 

\noindent \textbf{Funding.} This work was supported by the National Key R\&D Program of China (2023YFB3211200), by the National Natural Science Foundation of China (62250071, 62175127, 62375150, T2522035, 2476112), by the Tsinghua-Toyota Joint Research Fund, by the Quantum Science and Technology-National Science and Technology Major Project (2021ZD0300300) and by the HFNL Self-Deployed Project (ZB2025020300). 


\noindent\textbf{Author Contributions.} Z.C. led the experiments with assistance from Z.W., C.W., X.L., J.L. and C.Y.; Y.D., Y.W., Y.G., J.Y. and J.W. prepared the AlN chip; Z.C., C.Y., R.Z., X.X. and C.B. analyzed the results. Z.C. and C.B. led the writing of the paper with input from all the authors. The project was supervised by C.B.

\noindent \textbf{Competing Interests.} The authors declare no competing interests.

\noindent \textbf{Data Availability.} 
The data that support the plots within this paper and other findings are available in the paper.
}

\bibliography{main}
\end{document}